# Investigation of the Imperfect Interface at the Si/GaAs Heterojunction Formed by Ultra-Thin Oxide Interfacial Layer

*Md Nazmul Hasan, Yixiong Zheng, Junyu Lai, Edward Swinnich, Olivia Grace Licata, Mohadeseh A. Baboli, Baishakhi Mazumder, Parsian K. Mohseni, Jung-Hun Seo*[*]

Md Nazmul Hasan, Yixiong Zheng, Junyu Lai, Edward Swinnich, Olivia Grace Licata, Prof. Baishakhi Mazumder, Prof. Jung-Hun Seo*

Department of Materials Design and Innovation,
University at Buffalo, the State University of New York,
Buffalo, New York, U.S.A, 14260
*Email: junghuns@buffalo.edu

Mohadeseh A. Baboli, Prof. Parsian K. Mohseni

Microsystems Engineering Ph.D. Program,
NanoPower Research Laboratories,
Department of Electrical and Microelectronic Engineering
Rochester Institute of Technology, Rochester, NY 14623, USA

**The structure property of non-ideal Si/GaAs heterostructures that were integrated with the ultra-thin oxide (UO) tunneling interfacial layer has been systematically investigated. Si nanomembranes (NMs) were oxidized in different time periods prior to the hetero-integration process to create the non-ideal single-side passivated Si/GaAs heterostructure. The atomic level oxygen distribution and the degree of oxygen content in Si NM and GaAs were carefully investigated using the atom probe tomography (APT) and X-ray photoelectron spectroscopy (XPS) to trace the changes in chemical composition and reactional mechanism across the UO interface when the surface of Si NM was exposed to air for different period of time. The negatively induced charges at the UO layer caused**

**the oxygen diffusion to GaAs layer and formed the unwanted GaAs oxide layer. This native oxide stack noticeably degraded the thermal properties of the Si/GaAs heterostructure as Si NMs were more oxidized. This study revealed that the poor surface passivation on the one side of the heterointerface leads to a both-side oxidation, thus severely deteriorating the transport properties across the heterojunction that is formed with the UO layer.**

1. Introduction

Semiconductor heterojunctions are important building blocks that enable various modern electronics and optoelectronic applications such as heterojunction bipolar transistors (HBTs), high electron mobility transistors (HEMTs), light-emitting diodes (LEDs), and lasers. [1-5] These heterojunctions are routinely grown on lattice-matched or slightly lattice-mismatched substrates via molecular-beam epitaxy (MBE) or metalorganic vapor-phase epitaxy (MOVPE). The growth of heterostructures using these methods is largely restricted by their lattice parameters, thus, it is impossible to grow high-quality heterostructures that have different lattice constants or different crystal structures. [6, 7] Therefore, despite the advent of numerous new semiconductors in past decades, their implementation in the formation of heterostructures is limited or not possible by their material parameters. To overcome this issue, various studies have been performed; the first approach was to use a wafer bonding method to physically bond two different materials together. [8-10] However, this method suffers from a substantially thick interfacial layer (> 10 nm) due to the presence of native oxide layers from each substrate. It is also difficult to manage different thermal expansion rates and different degrees of surface roughness from both substrates. Surface-activated bonding (SAB) was introduced to deal with these issues by bonding two different wafers in a vacuum after plasma treatment for the native oxide removal. [11] SAB has been successfully applied to create numerous hetero-integration strategies of two completely dissimilar semiconductor wafers leading to various applications such as transistors, [12] photovoltaics, [13] and other applications. [14, 15] However, the presence of a few nanometer-thick interfacial layer is still a technological bottleneck that hinders ideal charge transport across the heterojunctions. The heterojunctions enabled by semiconductor grafting method via employing an ultra-thin oxide (UO) interface layer was the first approach to tackle the interfacial high density of states that existed between two dissimilar semiconductors. [16] This method uses a sub-nanometer-thick oxide layer (called the UO layer) to passivate both semiconductor surfaces after the native oxide removal and to facilitate charge transport via quantum tunneling. In recent years, various heterostructures with two largely dissimilar semiconductors and their device applications have been demonstrated. [3, 16, 19-21] While the first Ultra-thin Oxide (UO) heterointegration paper (Ref. 16) has successfully demonstrated the best possible Si/GaAs heterojunction with a well-suppressed native oxide growth both on Si and GaAs surfaces prepared under a controlled environment (such as the nitrogen-filled glovebox and the unified the UO growth system), this manuscript presents the systematical analysis when the one side of the heterointerface is not well passivated, namely, the Si nanomembrane (Si NM) side in this case. As is well-known, it is relatively easy to achieve the native oxide free GaAs surface with the UO layer, because some degree of native oxide on the GaAs surface can be effectively removed by purging additional trimethylaluminum (TMA) cycles in the atomic layer deposition (ALD) system prior to the $Al_2O_3$ deposition. On the other hand, the opposite side of the interface (i.e., the Si NM surface) is prone to get oxidized because it is inevitable that the Si NM surface is exposed to air after the releasing process and before the integration process onto the UO coated GaAs surface. Also, the Si oxidation is occurred during rinsing with deionized (DI) water and nitrogen ($N_2$) drying process, producing a thin unavoidable native oxide with the thickness range of 0.2 nm ~ 0.4 nm.[23] Although the best practice for the UO heterointegration is to keep to-be-bonded surfaces from any unwanted surface oxidation, it is possible that the one side of the UO interface can be oxidized when improper environmental or process conditions are made. With these reasons as the primary motivation, this work studies the importance of a proper surface passivation during the integration process by intentionally introducing the imperfect Si/GaAs interface to reveal the nanoscale degradation mechanism. Then, the atomic level oxygen distribution and the degree of oxygen content in Si NM and GaAs were carefully investigated using the atom probe tomography (APT) and X-ray photoelectron spectroscopy (XPS) to trace the changes in chemical composition and reactional mechanism across the UO interface when the surface of Si NM is exposed in air for different period of time. Interestingly, it was found that substantial amount of oxygen atoms were diffused from the Si native oxide layer and $Al_2O_3$ UO layer to the GaAs side and formed a 3 ~ 4 nm thick oxidized GaAs (GaAsO) layer. This thick native oxide layers also deteriorate the thermal property of the Si/GaAs structure. The thermal boundary conductance across the Si-$Al_2O_3$-GaAs structure was decreased by 12% when two different oxidized samples were measured.

**Result and Discussion**

**Figure 1(a)** shows a schematic illustration of the sample preparation steps to create the one side imperfect heterojunction using Si NM and GaAs. Details of the fabrication process can be found in the methods section. Also, various heterointegration with different material combinations using a UO method and related device fabrication method can be found in somewhere else. [16] In brief, the fabrication starts with the Si nanomembrane (NM) preparation. As shown in **Figure 1(a1)**, a top 50 nm ~ 70 nm Si device layer in insulator (SOI) substrate was partially doped by the boron implantation to create the p+/p- structure that helps to form Ohmic contact later. After the doping process, the entire top Si layer was patterned and immersed in hydrofluoric (HF) acid to wet etch the buried $SiO_2$ layer (**Figure 1(a2)**). Once the buried $SiO_2$ layer is completely etched, the released top Si layer (now called Si NMs) was pick up using an elastomeric polydimethylsiloxane (PDMS) stamp as shown in **Figure 1(a3)**. In this step, the picked Si NMs were exposed for two different time periods; 3 min and 10 mins prior to the following integration process. Meanwhile, a thin 0.7 nm of $Al_2O_3$ layer as a UO layer was deposited on n-/n+ GaAs substrate using atomic layer deposition (ALD) as shown in **Figure 1(b1)-(b3)**. Prior to the $Al_2O_3$ deposition, 20 cycles of pre-purging with TMA precursor were performed on the n-GaAs substrate to remove a native oxide from the GaAs surface. (**Figure 1(b2)**) The oxidized p+/p- Si NMs were gently transfer-printed onto the $Al_2O_3$ coated n-/n+ GaAs substrate using a PDMS stamp as shown in **Figure 1(c1)** followed by thermal annealing at 300 °C for 5 min using a rapid thermal annealing (RTA) system to enhance the bonding strength. Prior to the device characterization, we performed various material and structure-property investigations. The optical microscope image shown in **Figure 2(a)** confirms that no visual defect sites, mechanical cracks, or fractures were found on Si NM. The uniform pink color of the Si NM implies that the transfer-printed Si NM was bonded uniformly to the $Al_2O_3$ coated GaAs surface. **Figure 2(b)** shows a scanning electron microscopic (SEM) image of the transfer-printed Si NM on an $Al_2O_3$-coated GaAs substrate. The elemental composition analysis captured by energy-dispersive X-ray spectroscopy (EDS), shown in **Figure 2(c)**, indicates particle-free clean Si NM and $Al_2O_3$-coated GaAs substrate. Raman spectroscopy was carried out to investigate the strain from the Si/GaAs heterostructure that could have been induced during a transfer-printing or thermal annealing process. **Figure 2(d)** presents the Raman spectrum taken from the center of the Si/GaAs heterostructure. Raman modes at 292.6 cm$^{-1}$ for GaAs and at 520.8 cm$^{-1}$ for Si confirm that the Si/GaAs heterostructure does not experience any strain. However, the surface oxidation on Si NM was not detected by the Raman spectroscopy, thus it was not possible to quantify the degree of the surface oxidation on Si NM. We speculated that a few nanometer thin SiO$_x$ native oxide inside of the Si/GaAs heterostructure produced insufficient Raman signal. Therefore, the Si/GaAs heterointerface was analyzed using APT as shown in **Figure 2(e)**. APT provides atomic-level information on defects and chemistry not only from the bulk but also at the buried interface, which is critically useful in defining interfacial chemistry and impurities. [24] **Figure 2(e)** shows a three-dimensional atomic map of the Si/GaAs heterostructure revealing the presence of a thin Al$_2$O$_3$ layer sandwiched between Si (green dots) and GaAs (purple and blue dots). The atomic concentration of each element with respect to depth is shown as a 1D concentration profile in **Figure 2(f)**. Several important aspects of the Si/GaAs heterostructure can be observed here. Firstly, the abrupt reduction of As concentration and increase of Ga concentration near the Al$_2$O$_3$/GaAs were found. This concentration changes together with the long oxygen profile tail toward GaAs implies that oxygen was diffused from SiO$_x$ and oxygen vacancies in Al$_2$O$_3$ and formed GaO$_x$ by attracting Ga ions. At the same time, this chemical reaction pushes As ions inward, thus As ions become partially depleted in first 3~4 nm, while some of oxygen atoms were bonded with As (As-O). Previous studies on Al$_2$O$_3$/GaAs structure also reported the movement of As ions from the Al$_2$O$_3$/GaAs interface due to the oxidation.[25] Also, oxygen diffusivity in GaAs is high (much higher than the diffusivity in Si), thus oxygen diffusion toward GaAs can be occurred at a lower energy, [26] such as a low temperature annealing process, instead of a Si side. Due to these reasons, in this case, diffused oxygen ions are the primary agent that forces the passivated GaAs surface to oxidize.

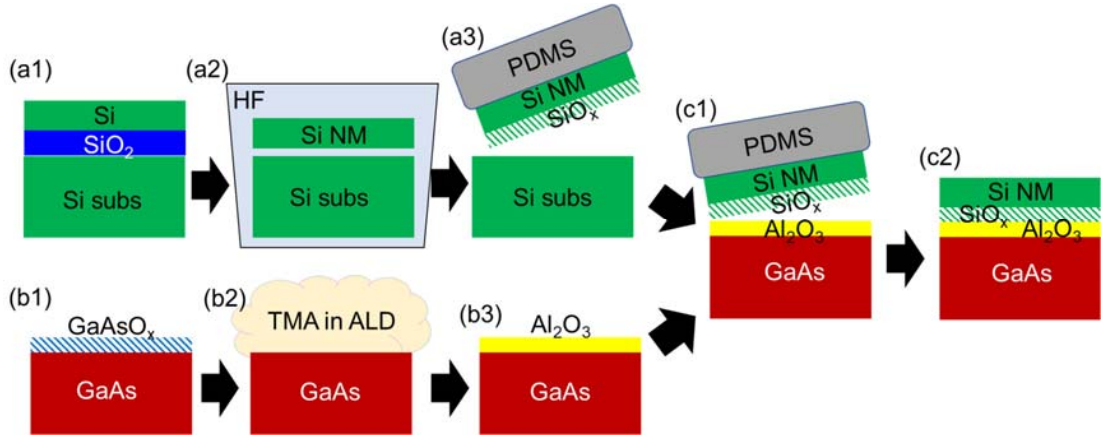

**Figure 1.** The schematic illustration of the sample preparation steps to create the one side imperfect heterojunction using a UO method, (a1)-(a3) a creation of Si NM from the SOI host wafer and expose Si NM to air for different time periods, (b1)-(b3) an $Al_2O_3$ UO layer deposition on GaAs epitaxy substrate with the native oxide removal process, (c1)-(c2) Si/GaAs heterogeneous integration via a micro-transfer printing method.

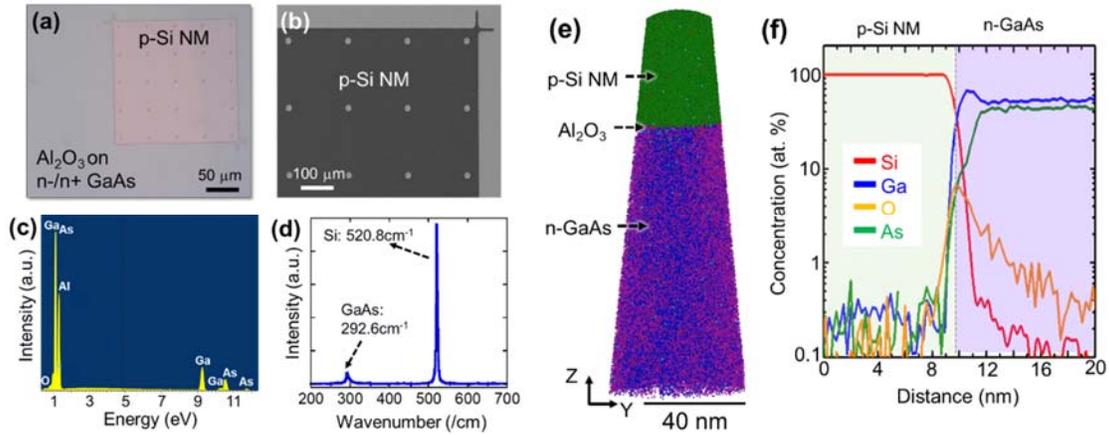

**Figure 2. (a)** Optical microscopic and **(b)** Scanning electron microscopic image of transferred Si NM on the Al$_2$O$_3$ coated GaAs substrate. **(c)** EDX spectrum taken from the Al$_2$O$_3$ coated GaAs substrate, **(d)** Raman spectrum taken from the transferred Si NM on the Al$_2$O$_3$ coated GaAs substrate, **(e)** 3D APT image taken from -/+ 50nm at the Si/GaAs heterointerface, **(f)** Atom concentration (Si, Ga, O, As) as a function of distance taken from APT.

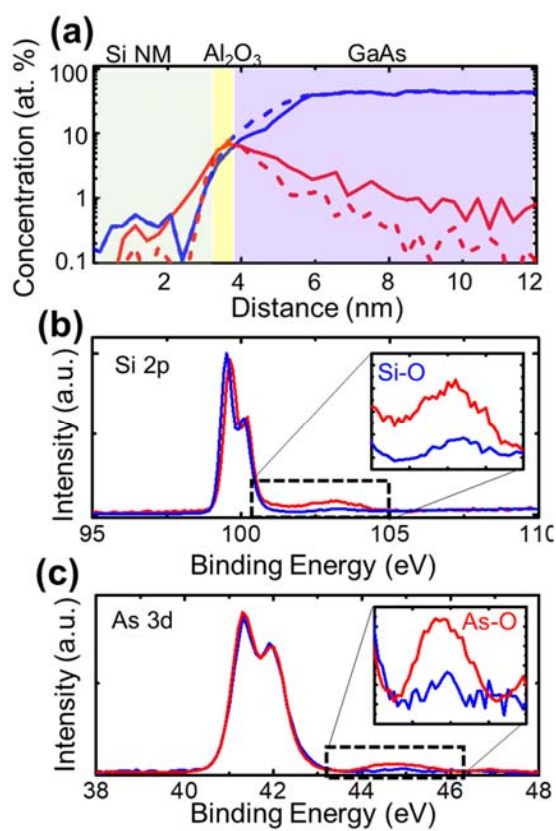

**Figure 3. (a)** Oxygen (red) and As (blue) concentration as a function of distance taken from APT. The dashed line and the solid line indicate samples with different Si NMs that were oxidized for 3 min and 10 min, respectively **(b), (c)** XPS spectra showing Si$_{2p}$ and As$_{3d}$ binding energies for samples with different Si NMs that were oxidized for 3 min (blue) and 10 min (red). Insets shows the zoom-in of the Si-O bond region and As-O bond region, respectively.

To investigate the detailed influence of oxygen in Si/GaAs imperfectly passivated heterostructure, **Figure 3** presents a more systematical analysis on the Si/GaAs heterointerfaces that have different Si oxidation time (i.e., 3 min and 10 min in the air). **Figure 3(a)** shows the oxygen profile from the APT profile for two differently oxidized samples. As expected, the oxygen diffusion length to GaAs estimated using a full-width half maximum (FWHM) is measured to be ~1 nm for 3 min oxidized Si NM sample, but this value increases to 3 ~ 4 nm for 10 min oxidized Si NM sample. The XPS spectra shown in **Figure 3(b) and (c)** also confirmed the presence of small oxygen atoms on both side of the heterointerface, namely the Si-O bond from the Si/Al$_2$O$_3$ side and the As-O bond from the Al$_2$O$_3$/GaAs side. Interestingly, the XPS spectra show that a strong Si-O bond and As-O bond were formed when Si NM was exposed to air for 10 min. However, when Si NM was exposed to air for 3 min, a nearly unnoticeably small Al-O bond signal was found, while Si-O bond was sufficiently strong. Such difference in the As-O bond is probably attributed to the development of native oxide on Si NM surface, because as is known, it takes about 15 ~ 20 min to complete the surface oxidation for Si, therefore, 3min and 10 min of oxidation on Si NM surface may produce a different amount (thickness) of incomplete native oxide. Regarding the detailed degradation mechanism associated with **Figure 3** will be discussed in the next paragraph that is related to **Figure 4**.

We used XPS not only to trace oxygen in Si/GaAs structure but also to investigate the energy band alignment between Si and GaAs with the Al$_2$O$_3$ UO interfacial layer.[27] The band offset measured from the XPS study explains that the Si and GaAs are connected with the Al$_2$O$_3$ UO interfacial layer and the interface defect densities ($D_{it}$) between Si and Al$_2$O$_3$ and Al$_2$O$_3$ and GaAs when Si NM surface is exposed to air in different time periods for 3 min and 10 min, respectively. To accurately quantify the band structure of the Si/GaAs heterostructure and their $D_{it}$ characteristics, the Si/GaAs heterostructure was divided into two sections: Si/Al$_2$O$_3$ and Al$_2$O$_3$/GaAs. Firstly, we prepared five samples with different structures as shown in **Figure S1** in the supplementary information to investigate the energy band offsets between Si/Al$_2$O$_3$ and GaAs/Al$_2$O$_3$ and performed XPS. Their spectra can be found in **Figure S2 and S3** in the supplmenetary information.

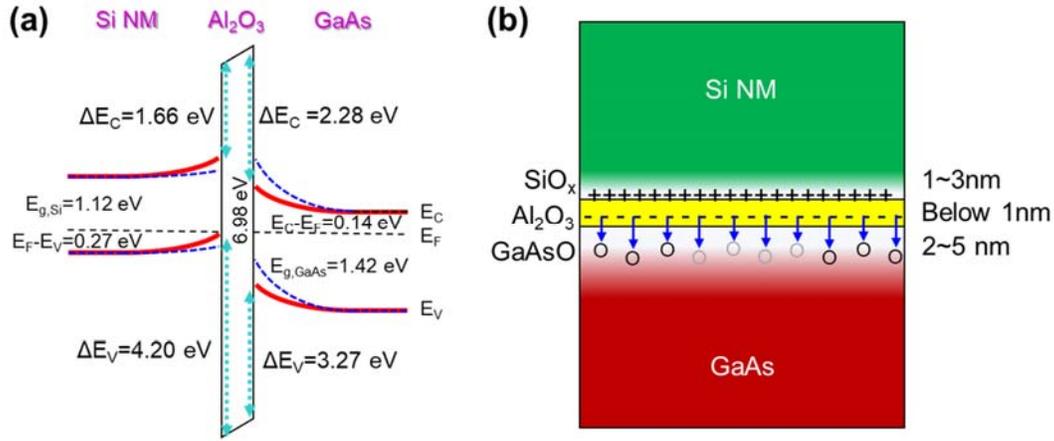

**Figure 4. (a)** Band diagram of Si/GaAs hetero junction under equilibrium condition, and **(b)** a cross-sectional illustration of the induced charges at the interface.

As described in the method section, the Si/GaAs heterostructure is consisted of four differently doped layers. The GaAs substrate has two epitaxy layers; the 500 nm n- layer (Si: $1 \times 10^{16}$ cm$^{-3}$) on top of a 500 nm n+ layer (Si: $5.8 \times 10^{18}$ cm$^{-3}$), and the Si NM has a 350 nm thick p+ layer (B: $1 \times 10^{20}$ cm$^{-3}$ cm$^{-3}$) on top of a 50-70 nm thick p-layer (B: $1 \times 10^{15}$ cm$^{-3}$) formed by an ion implantation. Therefore, the Si-GaAs heterointerface has a lightly doped p-n junction whose Fermi levels are expected to be 0.27 eV for Si and 0.14 eV for GaAs, respectively. In the ideal Si/GaAs heterojunction, the built-in potential (V$_{bi}$) would be 0.7 eV from {4.05 eV ($\chi_{Si}$) + 1.12 eV (E$_{g,Si}$) – 0.27 (E$_F$)} - {4.07 eV ($\chi_{GaAs}$) + 0.14 (E$_F$)}, however, the V$_{bi}$ from 1/C$^2$-V curve was measured to be 0.59 eV as shown in **Figure S4** which in turn a direct evidence of a possible Fermi-level pinning at the interface. With this, the band diagram under equilibrium condition was plotted based on the conduction band minimum (CBM) and valence band maximum (VBM) values from the XPS measurement. As is shown in **Figure 4(a)**, the CBM / VBM values between oxidized Si NM and Al$_2$O$_3$ are measured to be 1.95 eV / 3.91 eV and 1. 66 eV / 4.2 eV when the Si NM surface was oxidized for 3 min and 10 min, respectively. Unexpectedly, the CBM / VBM values between $Al_2O_3$ and GaAs were also changed from 2.17 eV / 3.39 eV and 2.28 eV / 3.28 eV, respectively, when the GaAs surface is bonded with Si NMs that were oxidized for 3 min and 10 min. This change is aligned with the XPS results in Figure 3, in that the longer oxidation in Si NM, the stronger oxide peaks were appeared on both sides. With these results, it is possible to draw a chemical reaction across the Si/GaAs heterointerface as shown in **Figure 4(b)**. In fact, the thicker a Si native oxide is, a stronger positive charges are presented which lead to a stronger stronger negative charges were induced at the $Al_2O_3$-GaAs interface. And this process is also related to the movement of CBM and VBM levels when Si NM is oxidized in different time periods. In other words, when the Si NM surface is oxidized, the VBM of Si NM tends to be increased, thus the hole accumulation becomes higher as depicted as red lines in Figure 4(a).

After the band diagram study, we prepared metal-oxide-semiconductor capacitor (MOSCAP) structures to evaluate the $D_{it}$ values using a C-V based conductance method. [28] The conductance method measures the equivalent parallel conductance ($G_p$) of a MOS capacitor as a function of bias voltage and frequency from C-V and capacitance-frequency (C-$f$) curves to extract the conductance (S) and eventually the $D_{it}$ values. **Figure S5(a-d)** in the supplementary document shows the C-V and G-V measurement at 100 kHz of Si/Al$_2$O$_3$ and GaAs/Al$_2$O$_3$ MOSCAPs with different Al$_2$O$_3$ thicknesses. The absence of a conductance peak in **Figure S5(b)** for the GaAs/Al$_2$O$_3$ structure is attributed to the interface trap loss caused by the series resistance. [29] The frequency dispersion that occurred at the accumulation region appeared from both Si and GaAs MOSCAPs, which suggests an interface defect near in the valence band (for p-Si) and conduction band (for n-GaAs). [30-32]

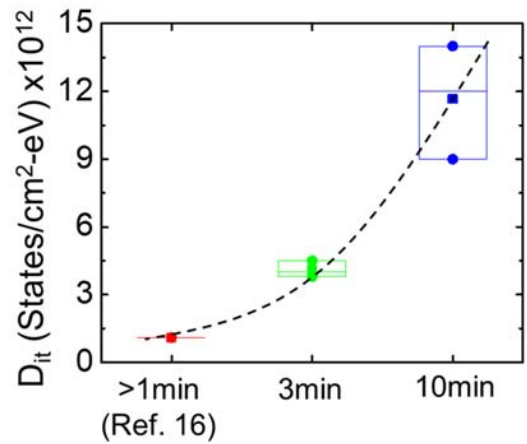

**Figure 5.** Measured interface defect density ($D_{it}$) taken from Si/GaAs heterostructures with Si NMs that were oxidized for 3 min and 10 min and compared these values with the ideal case from Ref. 16.

**Figure 5** presents the measured D$_{it}$ from Si/GaAs heterostructures with Si NMs that were oxidized for 3 min and 10 min. D$_{it}$ values were increased from $5.4 \times 10^{12}$ states/cm$^2$-eV to $1.2 \times 10^{13}$ states/cm$^2$-eV when Si NMs were oxidized for 3min and 10 min. Also, the variation of D$_{it}$ values become wider as the Si NMs were oxidized for a longer time. Compared with D$_{it}$ of the ideal Si/GaAs heterojunction shown in Ref 16, these D$_{it}$ values are severely higher and their distributions are wide. This trend can be explained by uneven native oxide thicknesses in Si NM and GaAs surfaces. Unlike thin-films grown by an epitaxy growth method, the oxidation chemical process is occured at a randam locations, thus a thickness of the native oxide layer can be varied upto 50 %. [33] The uneven oxide thicknesses in Si NM and GaAs lead to uneven charge distributions, thus lead to a wide variation in $D_{it}$. When widely distributed $D_{it}$ values were observed, it is a direct evidence of a presence of a native oxide, thus the transport property across the heterojunction can be degraded.

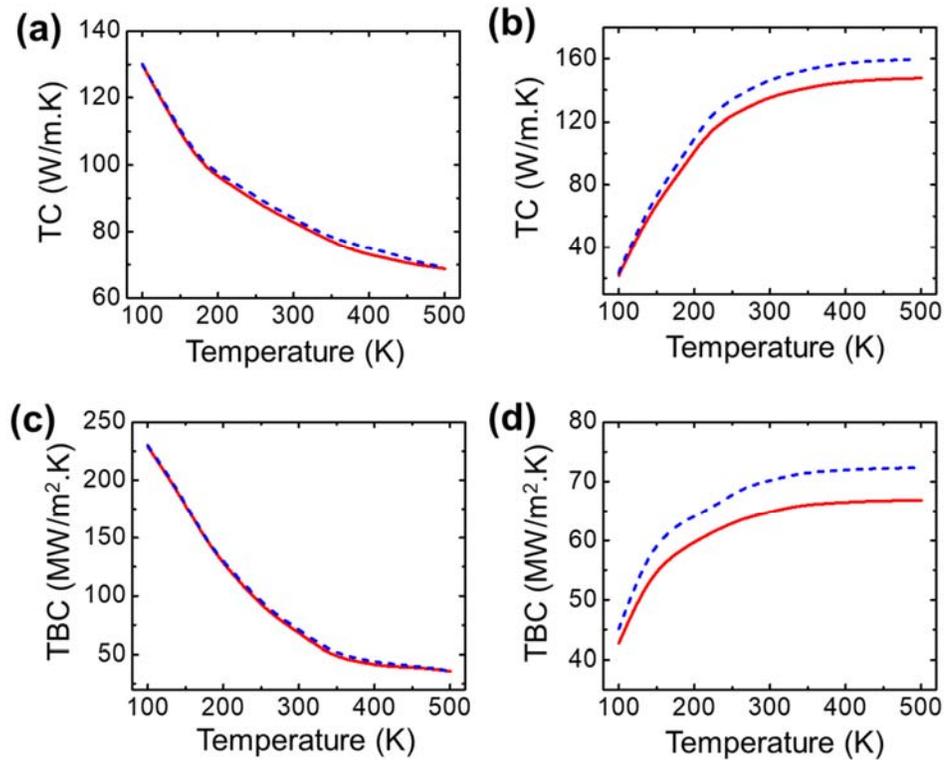

**Figure 6.** Measured thermal conductivity of **(a)** Si NM and **(b)** GaAs when they are bonded with Si NMs that were oxidized for 3 min (blue dashed line) and 10 min (red solid line) and thermal boundary conductance of **(c)** Si NM and **(d)** GaAs when they are bonded with Si NMs that were oxidized for 3 min (blue dashed line) and 10 min (red solid line) from 100 K to 500 K with a 40 K interval.

In fact, when a stack of native oxides is presented at the heterointerface, a phonon transport is also deteriorated, because native oxides typically have numbers of dangling bonds and effectively shorten a phonon mean free path. To investigate the changes of thermal properties in Si/GaAs heterostructures, a Raman thermometry was employed to extract thermal conductivity and thermal boundary conductance under a wide range of temperature conditions (from 100 K to 500 K with a 40 K interval). As shown in Ref 34, 35, and **Figure S6**, these thermal property values can be directly extracted from the Raman shifting as a function of different temperature conditions. [34, 35] As shown in **Figure 6(a) and (b)**, thermal conductivity (TC) of Si was measured to be 129.9 W/m·K and 128.2 W/m·K at 100 K for Si NM with 3 min and 10 min oxidation and gradually decreased to 82.9 W/m·K and 81.7 W/m·K at room temperature. TC of GaAs shows the similar trend; 228.9 W/m·K and 227.7 W/m·K at 100 K for Si NM with 3 min and 10 min oxidation and gradually decreased to 68.5 W/m·K and 67.4 W/m·K at room temperature. This trend can be explained by the phonon–phonon scattering rate being sensitive to temperature because fewer states are occupied. Unlike TC values, the thermal boundary conductance (TBC) value which measures the conductivity of phonon transport across the interface varies noticeably. TBC of Si/$Al_2O_3$ at room temperature was measured to be 147 MW/($m^2$·K) and 147 MW/($m^2$·K) for Si NM with 3 min and 10 min oxidation and $Al_2O_3$/GaAs to be 70.2 MW/($m^2$·K) and 70.2 MW/($m^2$·K) for the samples with 3 min and 10 min oxidized Si NMs. Such a different result in TC and TBC can be explained by the nature of TBC whose value is sensitive to the interface quality—namely a lower surface state at the interface reduces phonon transmission across the heterointerface, thus a higher TBC can be obtained. [36] On the other hand, TB represents the bulk thermal property which represents that the bulk properties of Si and GaAs remain unchanged. Therefore, the low TBC values imply that the phonon transport across the UO heterointerface was severely degraded by the stack of oxide layers.

**Conclusions**

In conclusion, the structure property of non-ideal Si/GaAs heterostructures that were integrated with the ultra-thin oxide (UO) tunneling interfacial layer has been systematically investigated. Si nanomembranes (NMs) were oxidized in different time periods prior to the hetero-integration process to create the non-ideal single-side passivated Si/GaAs heterostructure. The atomic level oxygen distribution and the degree of oxygen content in Si NM and GaAs were carefully investigated using the atom probe tomography (APT) and X-ray photoelectron spectroscopy (XPS) to trace the changes in chemical composition and reactional mechanism across the UO interface when the surface of Si NM was exposed to air for different period of time. The negatively induced charges at the UO layer caused the oxygen diffusion to GaAs layer and formed the unwanted GaAs oxide layer. This native oxide stack noticeably degraded the thermal properties of the Si/GaAs heterostructure as Si NMs were more oxidized. Such a thick oxide stack at the heterointerface causes the following issues: (i) generation of unwanted surface states from a presence of $SiO_x$ (silicon oxide) and $GaAsO_x$ (gallium oxide), thus resulting in poor Dit values and tunneling efficiency, (ii) Uneven $D_{it}$ variation across the surface due to the uneven oxide thicknesses in Si NM and GaAs surface, (iii) Unreliable electrical and thermal behavior due to a presence of the thick oxide stack. (iv) More importantly, this study revealed that the poor surface passivation on the one side of the heterointerface leads to a both-side oxidation, thus severely deteriorating the transport properties across the heterojunction that is formed with the UO layer.

**Experimental Section**

*p-Si/n-GaAs hetero-diode fabrication:* The fabrication started with the boron ion implantation onto the SOI wafer. A top 50 nm ~ 70 nm Si device layer in insulator (SOI) substrate was partially doped by the boron implantation to achieve $1 \times 10^{20}$ cm$^{-3}$ of boron concentration to create p+/p- structure that helps to form Ohmic contact. [37] A 500 μm x 500 μm sized Si NMs were patterned by a photolithography and defined using a reactive ion etcher (RIE) using SF$_6$/O$_2$ gases. A diluted HF: DI water (1:3) solution was used to etch a buried oxide layer of the SOI wafer for 7 hours in order to separate top p+/p- Si NM from the SOI wafer. The n+/n- GaAs wafers which have a 500 nm thick heavily Si-doped ($5.8 \times 10^{18}$ cm$^{-3}$) n+ layer and a 500 nm thick lightly Si-doped ($1 \times 10^{16}$ cm$^{-3}$) n- layer were grown by metalorganic vapor phase epitaxy (MOVPE) using an Aixtron 3×2-inch close coupled showerhead reactor. After the cleaning process, the Al$_2$O$_3$ UO interfacial layer was deposited using atomic layer deposition (ALD) system at 300 °C. The thickness of deposited Al$_2$O$_3$ was calibrated using ellipsometry, and it was found that the deposition rate was 1 angstrom per cycle. Prior to the Al$_2$O$_3$ deposition, 20 cycles of pre-purging with trimethylaluminum (TMA) precursor were performed on the n-GaAs substrate to remove a native oxide from the GaAs surface. The released Si NMs were picked up and transfer-printed onto a Al$_2$O$_3$ coated GaAs substrate using an elastomeric stamp (PDMS, Polydimethylsiloxane), followed by mild annealing (300 °C for 3 min) using rapid thermal annealing (RTA) to enhance the bonding strength. A cathode metal of Pd(30nm)/Ge(40nm)/Au(100nm) and an anode metal of Ti(10nm)/Au(150nm) were deposited on n-GaAs and p-Si side to finish the diode fabrication.

*X-ray photoelectron spectroscopy (XPS) measurement:* A Kratos Axis system with a monochromated Al-Kα (energy of 1486.7 eV) X-ray source was used to capture the different core level (CL) spectra. The CL binding energy and valence band binding energy spectra of Si (2p), Al (2p), and As (3d) were collected with a pass energy of 26 eV and an exit angle of 45°. The center of the peak width was considered to be a CL binding energy position of each spectrum. The emitted valence electrons from each sample determined from XPS measurement will define the band offsets (conduction and valence band) of $Al_2O_3$/p-Si and $Al_2O_3$/n-GaAs. The binding energy was calibrated by adjusting the carbon 1s CL peak at 285 eV for each sample surface.

***Atom probe tomography (APT) sample preparation:*** APT samples were prepared in an FEI Dual Beam 875 focused ion beam (FIB) through the standard wedge lift-out technique, [38] followed by annular milling to create sharp needle-like tips with an end radius less than 100 nm. APT acquisitions were carried out in a CAMECA, Inc. Local Electrode Atom Probe (LEAP®) 5000XR, equipped with a reflectron lens and ultraviolet (λ = 355 nm) laser pulsing capabilities. The APT analysis conditions consisted of 50 K base temperature, 5 pJ pulse energy, 0.3-2.0% detection rate, and 120-200 kHz pulse frequency. The Si-GaAs sample displayed in this work is a representative sample.

**Acknowledgments**
This work was supported by the National Science Foundation (Grant number: ECCS- 1809077) and partially by the seed grant by Research and Education in energy, Environment, and Water (RENEW) Institute at the University at Buffalo.